\documentclass[%
reprint,
showpacs,
showkeys,
amsmath,amssymb,
aip,
prl,
letter
]{revtex4-2}

\usepackage{graphicx} 
\usepackage{dcolumn} 
\usepackage{bm} 
\usepackage[page]{appendix}
\usepackage{amsmath}
\usepackage{float}
\usepackage{scalerel}
\usepackage{subfigure}
\usepackage{color}
\usepackage[hidelinks]{hyperref} 
\usepackage[normalem]{ulem}  

\begin{document}
\title{Progressive polymer deformation induced by polar activity and the influence of inertia}

\author{Andr\'{e}s R. Tejedor}
\affiliation{Department of Chemical Engineering, Universidad Polit\'{e}cnica de Madrid, Jos\'{e} Guti\'{e}rrez Abascal 2, 28006, Madrid, Spain}
\affiliation{Theoretical Physics of Living Matter, Institute of Biological Information Processing and Institute for Advanced Simulation, 
Forschungszentrum J\"ulich, 52425 J\"ulich, Germany}
\author{Jorge Ram\'{i}rez}
\email{jorge.ramirez@upm.es}
\affiliation{Department of Chemical Engineering, Universidad Polit\'{e}cnica de Madrid, Jos\'{e} Guti\'{e}rrez Abascal 2, 28006, Madrid, Spain}
\author{Marisol Ripoll}
\email{m.ripoll@fz-juelich.de}
\affiliation{Theoretical Physics of Living Matter, Institute for Advanced Simulation, 
Forschungszentrum J\"ulich, 52425 J\"ulich, Germany}
\date{\today}

\begin{abstract}
Polar activity is shown here to induce a progressive local deformation of linear polymer chains, making a clear distinction between head and tail, while the overall chain conformation gets more compact. 
This breakdown of self-similarity, provoked by the accumulated tension on the polymer backbone induced by the activity, is shown to occur both in the absence and presence of inertia, although it is stronger in the latter case. 
Other properties like the relaxation time and diffusion are also largely influenced by the presence of a polar activity.
\end{abstract}

\maketitle
Polymers in equilibrium are an archetypal example of fractal structure \textit{i.e.}, they are self-similar, so the shape of smaller sections replicates that of the overall macromolecule.
This fractal character has also been shown to persist in cases outside equilibrium, such as shear flow~\cite{leduc1999dynamics,winkler2010conformational}, or even in the presence of activity~\cite{winkler2020physics}. 
Intuitively, a polar activity applied on a linear polymer chain is expected to break up the head-tail symmetry, 
but previous investigations in this type of systems have not provided any indication of this particular behaviour~\cite{Bianco_2018,Anand_2018,winkler2020physics,Anand_2020,philipps2022tangentially}. In this sense, it is a fundamental question if there are circumstances in which polymers might not follow the self-similarity rules.

Self-propelling biopolymers are known to govern crucial biological functions within the cell, such as the replication of DNA by DNA-polymerase~\cite{alberts2017molecular},  transcription of RNA by RNA-polymerase~\cite{mejia2015trigger,guthold1999direct}, or the cellular motion promoted by actin microtubules~\cite{liu2011force,vutukuri2017rational,miosyn2007}.
Cytoskeletal molecular motors such as kinesin~\cite{1996kinesin,2004kinesin}, or  miosyn~\cite{miosyn2007,2002miosyn} induce microtubular polar propulsion, {\em i.e.} with a well-defined direction along the microtubule. 
Recently, it has been suggested that activity might influence the conformation and compartmentalization of polymer chains, aligning with the patterns observed in chromatin~\cite{Goychuk2023}.
The study of active polymers and filaments has gained much attention in recent years, particularly in the community of active matter~\cite{winkler2020physics,gompper20202020} where the overdamped approach is the most extended scenario. From the experimental perspective, different works have studied the dynamical and rheological properties of active polymers made of active colloids~\cite{dreyfus2005microscopic,hill2014colloidal,nishiguchi2018flagellar}, and also in the underdamped regime using polymer-like \textit{tubifex} worms~\cite{deblais2020rheology,deblais2020phase}. On the theoretical end, several models have been proposed to study the effect of active forces on linear polymers and implement the activity in different ways: using active Brownian particles~(ABPs)~\cite{Martin2019,Anand_2018,Anand_2020,kaiser2015does,natali2020local,farrell2012pattern} or Vicsek-like particles~\cite{paul2021motion,paul2022effects,paul2022activity} as monomers, with the activity  mimicking the interaction by an active bath~\cite{mousavi2021active,Martin2019,Harder2014,gandikota2022effective}, or the presence of explicit molecular motors~\cite{Vliegenthart_202Cactile}. 

\begin{figure}[b!]
	\centering
	\includegraphics[width=1.\columnwidth]{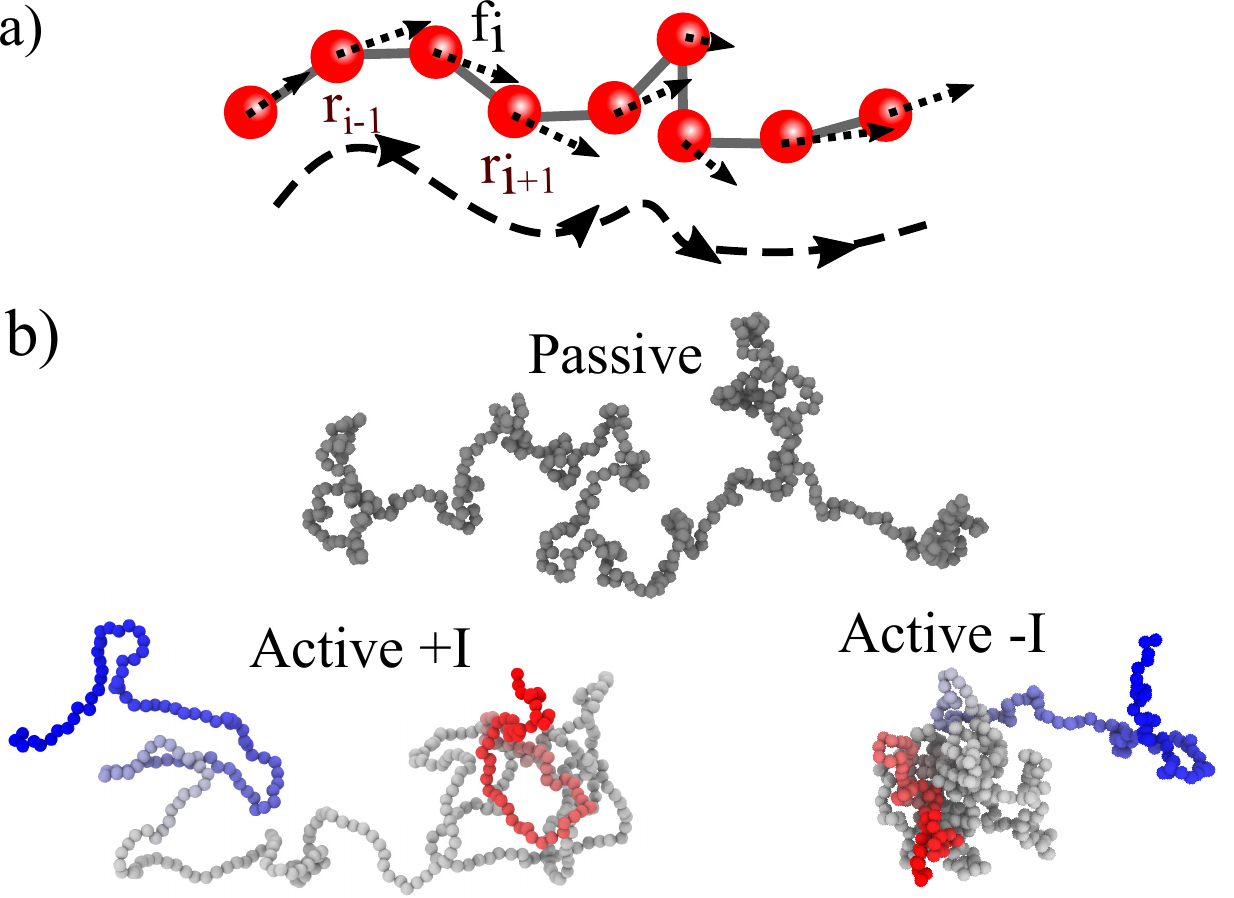}
	\caption{a)~Sketch of the polar active force implementation. Dotted lines are the forces $\mathbf{f}_{i}$ at each bead, and the dashed line is an eye-guide indicating the resulting force along the backbone. 
	b)~Snapshots of polymers with $N=400$ monomers, for the passive case as reference, and for the active cases with~(+I, Langevin dynamics) and without~(-I, Brownian dynamics) inertia, with monomeric P{\'e}clet numbers $\mathrm{Pe}_{m}=1$, and $\mathrm{Pe}_{m}=10$, respectively. Polymer heads and tails are colored in red and blue, respectively. Local stretching and global shrinking are noticiable in the~+I case, while head collapsing and tail stretching appear for both~+I and~-I cases. 
    }
	\label{fig:sketches}
\end{figure}
Recently, increasing interest has emerged in the dynamics of linear polymers subjected to polar and tangent activity, with works that study chains under dilute conditions~\cite{Bianco_2018,philipps2022tangentially}, ring polymers~\cite{Locatelli_2021,Philipps2022,Kumar2023,miranda2023self}, entangled polymers~\cite{Tejedor2019,Tejedor2020,Raquel2022}, translocation of chains through a pore~\cite{Tan2023}, and detachment from an attractive surface~\cite{Feng2023}.
Inertia has has been observed to induce qualitative behavioral changes in active matter systems, yet its exploration in polymer polar activity remains largely unexplored~\cite{jabbari23}.
In this letter, we investigate the conformation and dynamical properties of linear dilute polymers with and without 
inertia, and reveal that a progressive polymer deformation induced by polar activity is observed regardless of the presence of inertia. 

We consider here a linear flexible polymer model with $N$ beads connected by finite extensible non-linear elastic (FENE) springs~\cite{bird1987dynamics, kremer1990} with a mean bond length $b=0.97\sigma$. Non-consecutive beads interact via a repulsive WCA potential~\cite{Weeks_1971} that provides a bead steric diameter $\sigma$ (for details see SM~\cite{sm}). 
Polar activity acts on each monomer as a force $f_c$ tangentially to the chain contour, directed toward one of the ends, which is identified as the head. A central finite difference discretization of the derivative yields that, for intermediate monomers, the direction of the force is determined by the two nearest neighbors $\mathbf{f}_i=f_c\,(\mathbf{r}_{i+1}-\mathbf{r}_{i-1})/b$, 
while for end monomers, the force is applied in the bond direction (see Fig.~\ref{fig:sketches}a). Therefore, the force acting on each monomer depends on the local conformation, reaching a maximum modulus when the neighboring bonds are aligned and a minimum for antiparallel-like configurations. Here, hydrodynamic interactions are not considered, and the values of the activity are restricted so that the bond length distribution is not significantly affected (see Fig. S1). The inertial case~(+I) is simulated with underdamped Langevin dynamics, and simulations with no inertia~(-I) are performed with  fully overdamped Brownian dynamics (details in SM~\cite{sm}) equations of motion are integrated using LAMMPS \cite{Plimpton_1995, thompsonLAMMPSFlexibleSimulation2022}. Equivalence of passive dynamics is ensured by using the same monomeric friction coefficient both Brownian and Langevin simulations, here $\zeta_0=0.5$, unless otherwise specified.

The P\'{e}clet number of active polymers  typically relates to the active force to the surrounding thermal fluctuations. Here we employ  the {\em microscopic} or {\em monomeric} P\'{e}clet number, defined as  $ \mathrm{Pe}_{m}\equiv f_c b/(k_BT)$~\cite{Bianco_2018,Anand_2018,Anand_2020}, with $k_B$ the Boltzmann constant and $T$ the system temperature, which is responsible for local effects.  
On the other hand, the total force acting on the  polymer  $\mathbf{F}=\sum\mathbf{f}_i$ is proportional to the end-to-end vector and induces an instantaneous drift velocity~$V$ to the molecular center of mass (see SM~\cite{sm}). This velocity can be compared to the polymer equilibrium translational diffusion coefficient~$D_N$ and size~$R$, defining then the {\em polymeric} or {\em global} P\'{e}clet number~\cite{Isele_2015,philipps2022tangentially}, $\mathrm{Pe}_{g} \equiv V R/D_N$, which can be shown to be $\mathrm{Pe}_g = N\mathrm{Pe}_m$. 
 These two P\'{e}clet numbers are relevant to describe the activity effect on the chain conformations at different length-scales.

\begin{figure*}[htb!]
	\centering
	\includegraphics[width=\linewidth]{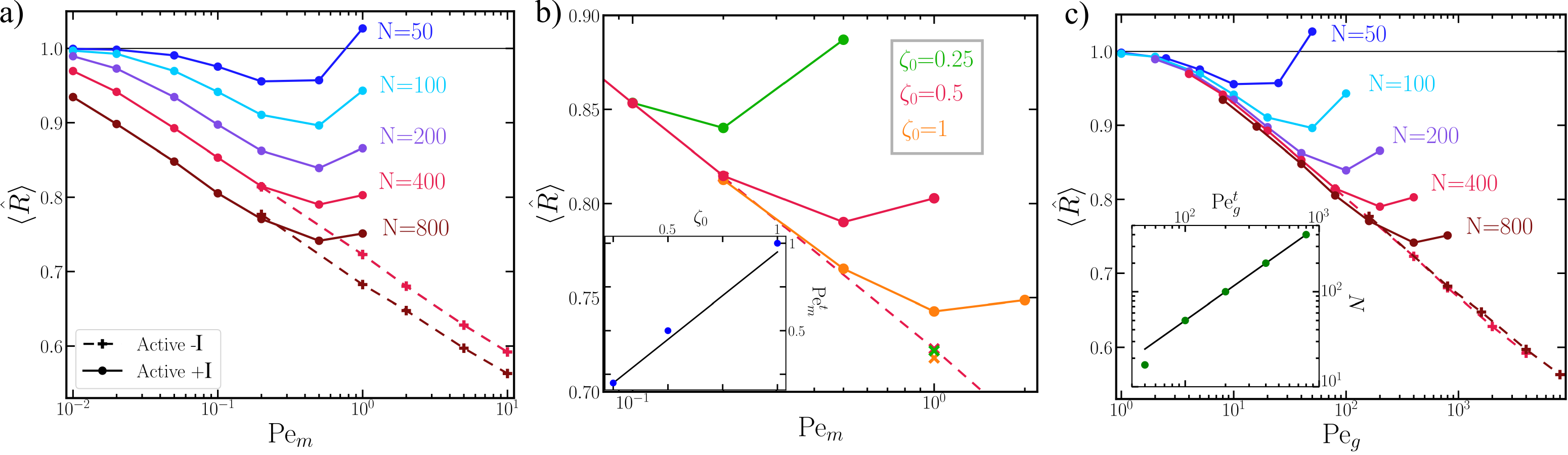}
	\caption{Polymer size for increasing activity, normalized by the values at equilibrium  for various polymer lengths $N$. Continuous lines and bullets correspond to the~+I case, while dashed lines and symbols correspond to the~-I case displayed as a function of ~$\mathrm{Pe}_{m}$ in~a),~b) and as a function of ~$\mathrm{Pe}_{g} = \mathrm{Pe}_{m} N$ in~c). 
    In a),~c)~simulations are performed at a fixed $\zeta_0=0.5$ value, both for +I and -I. 
    In b)~$N=400$ is fixed and $\zeta_0$ is varied. 
    Insets in b),~c)~display the crossover threshold values $\mathrm{Pe}_{m}^{t}$ and $\mathrm{Pe}_{g}^{t}$ respectively. 
	}\label{fig:pol-lenght}
\end{figure*}

Polymer conformations are strongly affected by the presence of polar activity, and this occurs both from global and local perspectives. From the global viewpoint, the overall polymer size, {\em i.e.} the end-to-end vector $\mathbf{R}$, is investigated for different polymer lengths $N$ and activities, as shown in Fig.~\ref{fig:pol-lenght}a, where $\langle\hat{x}\rangle$ denotes the average of the variable $x$ normalized by its value at equilibrium. 
At small $\mathrm{Pe}_{m}$ values, increasing the active force, or the polymer size, increases the chain shrinking independently of the presence of inertia. 
For large $\mathrm{Pe}_{m}$ values, polymers without inertia further shrink, while polymers with inertia change trend and swell.
This crossover behavior is characterized by a threshold $\mathrm{Pe}_{m}^{t}$ value, which Fig.~\ref{fig:pol-lenght}a shows to be independent of polymer length.
The strength of the inertia can be reduced by increasing the monomeric friction coefficient $\zeta_0$ of the Langevin thermostat (see SM~\cite{sm} for details) and, therefore, the value of $\mathrm{Pe}_{m}^{t}$ is expected to increase with $\zeta_0$.
Figure~\ref{fig:pol-lenght}b shows how increasing the friction coefficient in simulations with inertia, the minimum coil size is pushed to higher activities, roughly as $\mathrm{Pe}_{m}^t \simeq \zeta_0$, as shown in the Fig.~\ref{fig:pol-lenght}b inset. 
Previous studies of polar active polymers in the overdamped case showed monotonic decrease of the polymer size with $\mathrm{Pe}_m$, similar to our results~\cite{Anand_2018, Bianco_2018}.
The non-monotonic behavior of the polymer size with the applied activity, with swelling of the chains at large $\mathrm{Pe}_m$ values has also been observed for polar active polymers with inertia~\cite{jabbari23, Li2023}, where it was attributed to the activity of the head~\cite{Li2023}, and also in other Brownian dynamics studies of non-polar active polymers~\cite{Martin2019,Anand_2020}, where it was attributed to a rise in the local crowding of monomers at intermediate values of $\mathrm{Pe}_{m}$. 
Here, we relate this non-monotonic behavior to the combined effect of the activity of the head and the inertia, which prevents monomers from immediately following the path created by the head and the direction of the tangent active force. 

From a different perspective, the average coil size plotted as a function of the global P\'{e}clet number is shown in Fig.~\ref{fig:pol-lenght}c, where the decaying part of all the curves collapse into a universal line. 
This means that the relative global shrinking of the polymer chains is determined by the effect of the overall activity on the polymer, and not by the monomeric activity dictated by $\mathrm{Pe}_m$. In this view, the threshold value at which the inertia effect becomes relevant becomes polymer length dependent, as $\mathrm{Pe}_{g}^t \simeq N \zeta_0$, as shown in Fig.~\ref{fig:pol-lenght}c inset.
The universal part of the displayed shrinking is compatible with a logarithmic dependence of the polymer size with the P\'{e}clet number, as suggested by a phenomenological fitting function~\cite{Bianco_2018}, but in contrast to the analytical results of tangentially driven Rouse chains~\cite{philipps2022tangentially}, where excluded volume were not considered and for which the conformational properties resulted independent of the applied activity. 

\begin{figure}[t!]
	\centering
	\includegraphics[width=1.\linewidth]{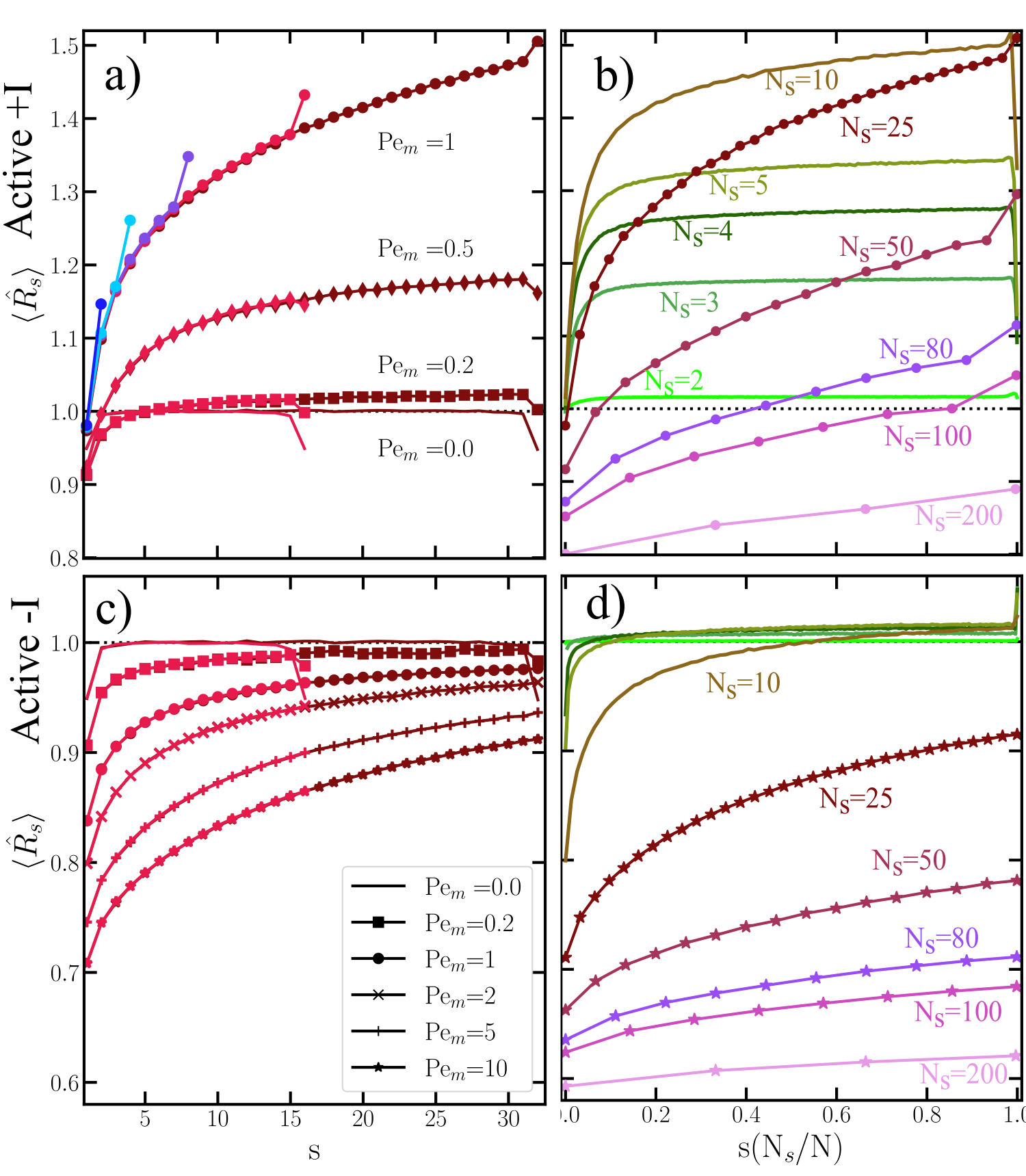}
	\caption{Mean size of chain strands  $\hat{R}_s$ along the chain contour, with $s$ the segment number and $s=1$ corresponding to the head. 
      a,b)~Simulations with Langenvin dynamics, c,d)~with Brownian dynamics. 
      a,c)~Segments with fixed $N_s=25$ monomers, for chains with different contour lengths $N$, and $\mathrm{Pe}_m$. 
      b,d)~Segments with variable $N_s$, for chains with $N=800$ monomers, 
      in b)~with $\mathrm{Pe}_m=1$, and in d)~with $\mathrm{Pe}_m=10$.} 
	\label{fig:segments}
\end{figure} 
Locally, polymer conformations are also significantly affected by polar activity, as can be seen by the non-homogeneous polymer deformation induced along the chain contour in the snapshots shown in Fig.~\ref{fig:sketches}b and in the movies in SM\cite{sm}.  
The reference passive case shows the characteristic self-similar shape of a random walk, whereas for the active case the tail segments are frequently more stretched than the head segments, causing the polymer to lose its self-similarity. 
Interestingly, this occurs both in the presence and the absence of inertia.  
To quantify the effect, we measure the extension of polymer segments along the contour. 
We calculate $\langle\hat{R}_s\rangle$, the mean end-to-end distance of segments of $N_s$ monomers along the backbone, normalized by the size of strands of the same length at equilibrium. Results of simulations in the underdamped limit are shown in  Figs.~\ref{fig:segments}a,b, and in the overdamped limit in Figs.~\ref{fig:segments}c,d.  Segments of length $N_s=25$ for different molecular weights and activities are analyzed in Figs.~\ref{fig:segments}a,c. In the passive case, only the dangling ends differ from the average.
In all active cases, the sizes of the polymer segments increase from the head ($s=1$) to the tail. Futhermore, for a constant activity, the strand sizes for different $N$ are identical, with the eventual exception of the very last segment. The extension of a given segment is therefore determined only by the number of monomers from the head to that particular segment,  and independent of the how many monomers away is the tail.
This is a consequence of the balance between the tension transmitted along the polymer backbone by the polar activity and the frictional resistance of the strand. The accumulated tension acting on monomer~$i$ is proportional to the strand size from the head to the monomer~$i$. This tension increases as the considered monomer is further away from the head, elongating the strands accordingly. As long as the chain from monomer~$i$ to the tail is large enough to have sufficiently high friction, the elongation of the strands is identical.
Consequently, the strands size along the chain follows a universal curve that depends only on $\mathrm{Pe}_m$.  The size of a strand of moderate size ($N_s=25$) is a local property, which explains why it depends on the monomeric and not on the global P{\'e}clet. 
The progressive segment size increase due to polar activity, could also be interpreted as an effective local stiffness, which increases from head to tail, and intensifies with longer polymer chains. The head is more flexible than the tail, resembling the shape of Gaussian chains with increasing stiffness along the contour~\cite{tejedor2022detailed}.

While the progressive stiffening of the polymer happens both in the presence and absence of inertia, there are important differences between both cases. 
With inertia, most chain segments expand (Fig. ~\ref{fig:segments}a), but the overall polymer size decreases (Fig. ~\ref{fig:pol-lenght}). This deviation from self-similarity suggests an inward-folding mechanism, as illustrated in Fig. 1b.
In 2D, where the motion is more restricted, this inwards folding induces the formation of spirals\cite{Isele_2015}. 
In constrast, without inertia nearly all segments contract in qualitative agreement with the overall chain, see Fig.~\ref{fig:segments}c. This indicates that the chains are adopting more globular conformations as can be seen in Fig.~\ref{fig:sketches}c.
Quantitatively, the effect of the activity on the polymer segments is much larger in the presence of inertia, and in order to get a comparable local deformation in the overdamped case, activities need to be typically an order of magnitude larger than in the underdamped case. 

The degree of  segment deformation changes qualitatively with the chosen size of the local strands $N_s$. 
In Fig.~\ref{fig:segments}b,d, the mean size of segments along the contour for chains of $N=800$ with different $N_s$ values are shown. In the underdamped case, the degree of stretching is a non-monotonic function of $N_s$, reaching a maximum at $N_s=10$, from which the stretching decreases to the point that, for $N_s\ge 100$, the segments get smaller than at equilibrium, in qualitative agreement with the behavior of the whole chain. 
In contrast, in the overdamped case, the segments get monotonically more compressed as $N_s$ increases, as shown in Fig.~\ref{fig:segments}d. 
In conclusion, inertia changes the qualitative behavior of active polar chains and is more effective at short length-scales. We expect that the range of strand lengths for which the local deformation can be observed experimentally must start at a few times larger than the Kuhn length until a few fractions of the total molecular weight.

\begin{figure*}[hbt!]
	\centering
	\includegraphics[width=2.0\columnwidth]{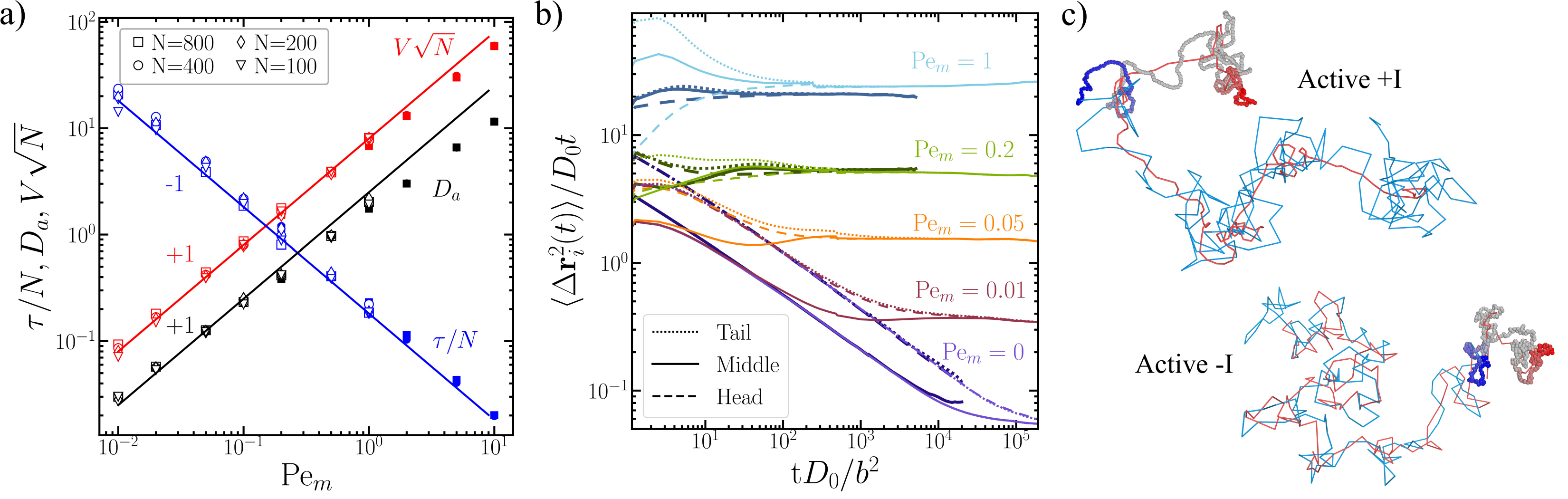}
	\caption{a)~Effective average polymer velocity, diffusion coefficient, and relaxation time for different polymer lengths and monomeric P{\'e}clet numbers. 
    Empty symbols correspond to +I simulations, and solid symbols to -I, all obtained from the fits to the averaged center-of-mass MSD.  
   Solid lines are a guide to the eye, indicating power laws with exponents $\pm 1$. 
    b)~Tail, middle, and head monomers MSD for $N=400$, and various $\mathrm{Pe}_{m}$ values. 
    Lighter lines correspond to +I results, darker lines show the -I counterpart. 
    The time axis is normalized by $D_0/b^2$, where $D_0 = k_BT/\zeta_0$ is the monomeric diffusion coefficient.
	c)~Snapshots of typical polymer configurations for $N=400$, together with the trajectories of the head (red) and the tail (blue) monomers, over a fixed period of time, for $\mathrm{Pe}_{m}$=1.
 }
	\label{fig:dynamics}
\end{figure*}
Polymer dynamics is also influenced by activity from both a global and a local perspectives.  
Similar to other active non-polar~\cite{Anand_2020,Martin2019} and polar polymers~\cite{Isele_2015,Bianco_2018}, the polar active polymer center-of-mass enhanced diffusion coefficient is independent of the molecular weight $N$ as shown in Fig.~\ref{fig:dynamics}a.  
A scaling argument is here used to rationalize such dependence. 
The total force acting on the center of mass $\mathbf{F}$ is proportional to the end-to-end vector, $\mathbf{R}$, oriented from tail to head, namely $\mathbf{F} = \sum \mathbf{f_i} \propto \mathrm{Pe}_{m} \mathbf{R}$, and it can be assumed that $|\mathbf{R}| \propto N^{0.5}$. Thus, the center of mass velocity is given by $V=F/N\zeta_0\propto\mathrm{Pe}_{m}/(N^{0.5}\zeta_0)$.
The orientational relaxation time $\tau$ is estimated to be the time it takes for the tail monomer, moving with speed $\mathrm{Pe}_{m}/\zeta_0$, to travel along a distance proportional to the polymer contour length and therefore to $N$, such that $\tau \propto N \zeta_0/\mathrm{Pe}_{m}$. From the diffusion coefficient of an ABP~\cite{Howse_2007,Cates2012,Cates2012,Zottl_2016}, $D_a \simeq V^2\tau/3$ 
we obtain the dependence $D_a \propto \mathrm{Pe}_{m}/\zeta_0$. 
The above scaling arguments are convincingly confirmed by the simulation data for all polymer lengths and activities studied (Fig.~\ref{fig:dynamics}a). 

To analyze local dynamics, the normalized MSD of selected monomers is displayed in Fig.~\ref{fig:dynamics}b. 
For small values of the activity, a long subdiffusive regime with Rouse scaling ($\propto t^{1/2}$) can be observed for all monomers, similar to the passive case~\cite{doi1988}, the central monomer being the slowest and with identical dynamics for the tail and head monomers, with no significant differences between the cases with and without inertia. 
For larger values of activity, the head bead starts to show a brief superdiffusive behavior, becoming the slowest bead at short times. Other monomers are progressively faster, and the tail becomes the fastest monomer with subdiffusive regime. 
This is illustrated by a representative polymer configuration in Fig.~\ref{fig:dynamics}c, together with the trajectories of the head and tail monomers. The path traveled by the tail is clearly much longer than that of the head, although at long times the mean distance covered by both ends is necessarily the same. 
The tail loosely tracks the path of the head, exhibiting lateral fluctuations that grow with activity. 
This behavior is illustrated in the monomeric MSD shown in Fig. 4b (see movies in SM~\cite{sm}).

Our results show how, not only global, but also local dynamical and conformational polymer properties are crucially altered by the presence of polar activity, both in presence and in absence of inertial effects. 
By construction, the polar activity introduces a tension along the chain contour 
that increases from head to tail, originating an induced progressive stiffness that breaks down the polymer self-similarity, which has not been observed in previous studies of this type of systems.  
It is also clear that inertia, which prevents monomers from moving immediately in the direction pointed by the tangent active force, increases this tension and further enhances the progressive polymer stiffness.
These are key elements for the full understanding of the properties of polymeric systems with polar activity.
Further experimental and theoretical investigations of polymers subjected to polar activity are expected to contribute to the design of drug delivery carriers with high mobility, where the induced chain asymmetry can improve the transport of different chemicals in dilution. 

\begin{acknowledgements}
A.T. acknowledges funding from EMBO scientific exchange grants and EBSA bursary, J. R. acknowledges funding from the Spanish Ministry of
Economy and Competitivity (PID2019-105898GA-C22)
and the Madrid Government (Comunidad de Madrid-Spain) under the Multiannual Agreement with Universidad Polit{\'e}cnica de Madrid in the line Excellence Programme
for University Professors, in the context of the
V PRICIT (Regional Programme of Research and Technological
Innovation).
The authors gratefully acknowledge the computing time granted by the JARA Vergabegremium and provided on the JARA Partition part of the supercomputer JURECA at Forschungszentrum J\"ulich~\cite{Jureca2016}, as well as the computer resources and technical assistance provided by the Centro de Supercomputaci\'{o}n y Visualizaci\'{o}n de Madrid (CeSViMa).
\end{acknowledgements}

 
%

\end{document}